\documentclass[usenatbib]{mn2e}

\usepackage{psfig,morefloats,url}
\def\lapp{\ifmmode\stackrel{<}{_{\sim}}\else$\stackrel{<}{_{\sim}}$\fi} 
\def\gapp{\ifmmode\stackrel{>}{_{\sim}}\else$\stackrel{>}{_{\sim}}$\fi}

\footnotesize
\newdimen\digitwidth    
\setbox0=\hbox{\rm0}
\digitwidth=\wd0
\catcode`!=\active
\def!{\kern\digitwidth}
\normalsize

\title[Parkes deep multibeam survey pulsars]
{Timing of pulsars found in a deep Parkes multibeam survey}
\author[D.~R.~Lorimer et al.]
{D.~R.~Lorimer$^{1,2,3}$,\thanks{Email: Duncan.Lorimer@mail.wvu.edu}
F.~Camilo$^{4,5}$ and
M.~A. McLaughlin$^{1,3}$
\\
$^1$ Department of Physics, West Virginia University, PO~Box~6315, Morgantown,
WV~26506, USA\\
$^2$ National Radio Astronomy Observatory, PO~Box~2, Green Bank, WV~24944, USA\\
$^3$ Astrophysics, University of Oxford, Denys Wilkinson Building, Keble Road, Oxford OX1 3RH\\
$^4$ Columbia Astrophysics Laboratory, Columbia University, 550 West 
120th Street, New York, NY 10027, USA \\
$^5$ National Aeronomy and Ionospheric Center, Arecibo Observatory,
HC~3 Box 53995, Arecibo, PR 00612, USA}
\date{Accepted for publication in MNRAS}
\begin{document}

\maketitle
\newcommand{\setthebls}{
}

\setthebls

\begin{abstract} 
We have carried out a sensitive radio pulsar survey along the northern
Galactic plane ($50^{\circ} < l < 60^{\circ}$ and $|b| \lapp
2^{\circ}$) using the Parkes 20-cm multibeam system. We observed each
position for 70-min on two separate epochs. Our analyses to date have
so far resulted in the detection of 32 pulsars, of which 17 were
previously unknown. Here we summarize the observations and analysis
and present the timing observations of 11 pulsars and discovery
parameters for a further 6 pulsars. We also present a timing solution
for the 166-ms bursting pulsar, PSR~J1938+2213, previously discovered
during an Arecibo drift-scan survey. Our survey data for this pulsar
show that the emission can be described by a steady pulse component
with bursting emission, which lasts for typically 20--25 pulse
periods, superposed.  Other new discoveries are the young 80.1-ms
pulsar PSR~J1935+2025 which exhibits a significant amount of 
unmodeled low-frequency noise in its timing residuals,
and the 4.2-ms pulsar PSR~J1935+1726 which is in a low-mass binary system
with a 90.7-day circular orbit.
\end{abstract}

\begin{keywords}
pulsars: general --- pulsars: searches --- pulsars: timing --- stars: neutron
\end{keywords}

\section{INTRODUCTION}

With over 800 new pulsar discoveries, primarily in the region defined
by Galactic coordinates $-100^{\circ} < l < 50^{\circ}$ and
$|b|<5^{\circ}$, the Parkes Multibeam Pulsar Survey (Manchester et
al.~2001) is the most successful large-scale search for pulsars to
date.  The survey, with 35-minute pointings, has been a fantastic
probe of the Galactic pulsar population and has resulted in the
discovery of numerous interesting individual systems (see, e.g., Lyne
2009 for a review).  Motivated by the high density of young pulsars in
the range $|b|<1^{\circ}$ (see, e.g., Camilo 2004), in 2004--2005 we
extended the coverage by surveying \nocite{cam04} the most northerly
Galactic longitudes possible from Parkes using 70-minute pointings in
the range $50^{\circ} < l < 60^{\circ}$ and $|b| \lapp 2^{\circ}$.
This region is also being covered by the Pulsar Arecibo L-band Feed
Array survey \citep{cfl+06}, and the Effelsberg High Time Resolution
Universe survey (Barr et al., submitted).  Our analyses to date have
resulted in the discovery of 17 new pulsars and a further 15 that were
previously known.  In \S\ref{sec:timing} we summarize the survey
observations and analyses so far, and present the timing parameters
and pulse profiles from observations of the newly discovered pulsars.
We discuss these results in \S\ref{sec:discssion} 
and present our conclusions in \S\ref{sec:conclusions}.

\begin{table*} 
\caption{Astrometric and spin parameters for the 18 pulsars 
described in this paper.  For the 12 pulsars for which timing
observations have been carried out, we also give the MJD of the epoch
used for period determination, the number of individual
TOAs ($N$) included in
the timing solution, the MJD range covered and the root-mean-square 
 (RMS) of the post-fit
timing residuals.  Figures in parentheses represent 1-$\sigma$
uncertainties in the least significant digit(s) as reported by \textsc{tempo}.}
\label{tb:posn}
\begin{center}\begin{footnotesize}
\begin{tabular}{lllllllll}
\hline
PSR J    &R.A. (J2000) & Dec. (J2000) & Period, $P$ & $\dot{P}$ & Epoch &
$N$ & Data span & RMS       \\
         &             &              &  (ms)        &(10$^{-15}$)& (MJD)&
              & (MJD)     &  (ms)     \\
\hline 
1919+1645&19:19:09.243(2)& 16:45:22.72(8)& 562.789986697(3) & 0.2161(4) &53560& 68 & 53286--53834 & 2.1 \\
1921+1544&19:21:46.4723(9)& 15:44:17.46(2)&143.5756814094(2)& 0.98036(2)&53632& 49 & 53347--53916 & 0.6 \\
1922+1733&19:22:53.2180(7)& 17:33:23.47(2)&236.1708772317(3)&13.38394(4)&53595& 108 & 53273--53916 & 1.2\\
1924+1639&19:24:03.1115(5)& 16:39:40.756(9)&158.0429177826(2)& 2.56384(1)&53720& 123 & 53272--53916 & 1.1\\
1925+19  &19:25:26(28)   & 19:04(7)      &1916.353(6)        &           &54346&     &              &     \\ \\ 
1926+2016&19:26:18.064(1)& 20:16:01.13(3)&299.0717782966(7) & 3.50084(6) &53595& 72 & 53271--53916 & 1.6  \\
1928+1923&19:28:05.163(5) & 19:23:31.32(8) &817.329808312(6)  & 6.3495(7) &53640& 125 & 53360--53916 & 2.2 \\
1929+16   &19:29:18(28)      & 16:21(7)  &529.681(2)       &           &53480&     &              &     \\
1929+19   &19:29:32(28)      & 19:05(7)  &339.2154(4)      &           &53363&     &              &     \\
1929+1955&19:29:17.5713(6)& 19:55:07.91(1)&257.8319398983(3)& 2.55530(3)&53595& 308 & 53271--53916 & 1.0 \\ \\
1929+2121&19:29:04.239(2)& 21:21:22.68(3)&723.598503258(2) & 2.1386(2) &53635& 91 & 53355--53916 & 0.3  \\
1930+17   &19:30:44(28)  & 17:25(7)      &1609.6903(7)     &           &53443&     &              &     \\
1931+1952&19:31:55.864(3)& 19:52:11.5(1) &501.123112694(4) & 0.1008(6)  &53560& 29 & 53285--53834 & 0.8  \\
1935+1726&19:35:03.948(3)& 17:26:28.46(3)&  4.200101791882(7)&$<0.00001$&55314&  23& 55112--55514& 0.049 \\
1935+2025&19:35:41.941(3)& 20:25:40.1(3)& 80.118133455(6) &60.7579(1)&53460& 109 & 53271--53650 & 0.3 \\ \\
1936+21   &19:36:29(28)      & 21:12(7)      &642.932(2)          &           &53281     &     &              &     \\
1938+20   &19:38:12(28)  & 20:10(7)      &687.0804(1)      &           &53286&     &              &     \\
1938+2213&19:38:14.180(3)& 22:13:12.68(4)&166.1155731566(6) &42.4423(5)  &53500& 79 & 53346--53651 & 0.8  \\
\hline 
\end{tabular}\end{footnotesize}\end{center}
\end{table*}

\begin{figure*} 
\centerline{\psfig{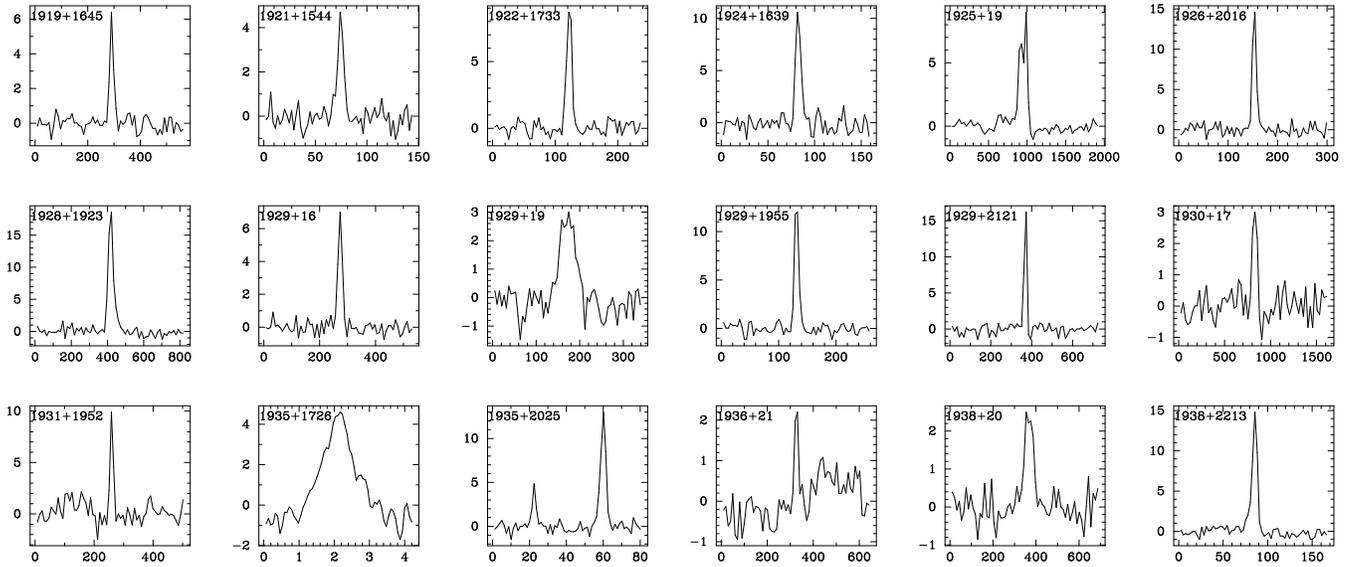}} 
\caption{Integrated pulse profiles for all 17 pulsars discovered in
the survey, as well as PSR~J1938+2213. The horizontal axis shows
pulse phase in ms and the vertical axis is flux density in mJy.
These data were obtained by 
folding one of the survey observations for each pulsar.}
\label{fg:prf}
\end{figure*}

\section{DISCOVERY AND TIMING OBSERVATIONS}\label{sec:timing}

We summarize the survey observations and data analysis procedures
which led to the discovery of the pulsars presented in this paper. 
The survey observations were conducted between
2004 September 22 and 2005 April 20 using the 20-cm multibeam receiver
at the Parkes 64-m telescope. A total of 120 multibeam pointings were
observed in the survey region approximately defined by Galactic
longitudes in the range $50^{\circ} \lapp l \lapp 60^{\circ}$ and
Galactic latitudes in the range $|b| \lapp 2^{\circ}$. Each sky
position was observed for 70 minutes at two different epochs, i.e.~the
entire survey region was covered twice.  Radio-frequency signals from
each beam were acquired using the analog filterbank system described
by Manchester et al.~(2001). The incoming data were sampled every
125~$\mu$s with 1-bit precision over 96 adjacent frequency channels
each of width 3 MHz.  The survey integration time and sampling rate on
each pass was twice that of the original multibeam survey (Manchester
et al.~2001).

An initial analysis of the data was performed shortly after collection
at Parkes on local computers using a pipeline based on the {\sc sigproc} 
processing software\footnote{http://sigproc.sourceforge.net}. To speed
up the overall processing time, the data were downsampled by adding
eight adjacent time samples together to form 96 channel filterbank
files with 1 ms time resolution. The effects of the interstellar
medium were compensated for by de-dispersing data from each telescope
beam at 496 different trial values of dispersion measure (DM) spanning
the range 1.1 and 4931~cm$^{-3}$~pc.  The de-dispersed time series
were then subjected to a standard Fourier analysis to search for
periodic signals \citep{lk05} and a search for dispersed single pulses
\citep{mll+06}.  Pulsar candidates were selected by eye from
inspection of diagnostic plots showing the properties of each
candidate. The two passes of the survey region allowed us to confirm
the existence of each pulsar reported in this paper by inspecting the
corresponding output of the second observation. This initial
processing resulted in the discovery of 12 pulsars. A second pass of
the data using the same pipeline, but with full time resolution,
resulted in the discovery of a further five pulsars including the
millisecond pulsar J1935+1726.  For all pulsars found so far, the data
have been folded in turn at two and three times the tabulated periods
to confirm that they represent the fundamental periods of the pulsars,
rather than a harmonic.

Timing solutions have so far been obtained for 11 of the newly
discovered pulsars as well as PSR~J1938+2213 which was previously
discovered (but not further studied to our knowledge) during a
drift-scan survey with the Arecibo telescope \citep{cha03}. These have
been obtained from the analysis of a series of observations carried
out with the Parkes telescope and Green Bank Telescope.  At Parkes, we
used the analogue filterbanks to acquire the data in an identical
fashion to that described by \citet{mlc+01} and \citet{mhl+02}
respectively.  The Green Bank timing made use of the SPIGOT and (for
J1935+1726) GUPPI data acquisition systems described by \citet{kel+05}
and \citet{drd+08} respectively. For each pulsar, pulse times of
arrival (TOAs) were determined from the individual observations using
standard techniques \citep[see, e.g.,][]{lk05} implemented in the
\textsc{presto} software
package\footnote{\url{http://www.cv.nrao.edu/~sransom/presto}}. A
model containing the spin, astrometric and (if necessary) any binary
parameters was fitted to the TOAs using the \textsc{tempo} timing
package\footnote{\url{http://tempo.sourceforge.net}}.  The basic
astrometric and spin information from this analysis is provided in
Table~\ref{tb:posn}.  For the six pulsars which have so far not been
timed, only their discovery information (i.e.~approximate position,
barycentric period and dispersion measure) are provided. Dispersion
measures and their quoted uncertainties are derived from the discovery
observations.  Specifically, the uncertainties are calculated by
requiring that the pulse must drift by less than one phase bin due to
any incorrect dispersion across the 288~MHz band.

\begin{table}
\caption{Orbital parameters for PSR~J1935+1726 obtained using
the ``ELL1'' binary model (Lange et al.~2001). Figures in parentheses
represent 1-$\sigma$ uncertainties in the least significant digit(s)
and are the nominal uncertainties reported by \textsc{tempo}.
We also list the equivalent orbital eccentricity, as derived from
the Laplace-Lagrange parameters (see Lange et al.~2001 for details).}
\begin{tabular}{ll} \hline
Orbital period (d)                          & 90.76389(2)\\
Projected semi-major axis of orbit (lt sec) & 31.97423(6)\\
First Laplace-Lagrange parameter            & 0.000031(4)\\
Second Laplace-Lagrange parameter           & 0.000173(4)\\
Epoch of ascending node (MJD)               & 54616.7206(1)\\
Orbital eccentricity                        & 0.000176(4)\\
\hline
\end{tabular} 
\label{tb:binary}
\end{table}

Our follow-up observations of PSR~J1935+1726 showed changes in the
barycentric spin period from epoch to epoch that were consistent with
the Doppler modulation due to binary motion of the pulsar.  A fit to
the barycentric periods revealed a 90.7~day period for a circular
orbit with a semi-major axis of 31.97~light seconds.  Once this orbit
was included in the timing analysis, we were able to obtain a fully
coherent fit using \textsc{tempo}. As is now common practice for
low-eccentricity binary systems, we used the ``ELL1'' binary model
using the Laplace--Lagrange parameterization of the eccentricity
\citep{lcw+01} in this fit. The resulting parameters are given in
Table~\ref{tb:binary} and discussed below in \S~\ref{sec:1935+1726}.

In addition to the timing analyses, we also carried out preliminary
measurements of pulse width and flux densities for each pulsar using
the survey data. The results of these measurements are given in Table
2.  Pulse widths were measured at 50\% and 10\% of the main pulse
($W_{50}$ and $W_{10}$ respectively) as well as the equivalent width
($W_{\rm eq}$) which is defined to be the width of a top-hat pulse
with the same area and peak flux density as the observed profile.  The
millijansky scale for the flux density calculations was established
from the off-pulse RMS using the radiometer equation ({\bf e.g.,}
Lorimer \& Kramer 2005) and adopting the Parkes multibeam system
parameters from Manchester et al.~(2001). For each profile, we
quote the peak flux density ($S_{\rm peak}$) and
calculated the mean flux density ($S_{\rm mean}$) by integrating
each profile and dividing by the number of bins. To account for the
position offset $r$ between the search-mode observation and the timing
position, we assumed a Gaussian beam with half-power beamwidth $w$ and
multiplied the flux densities by the factor $\exp(2.77r^2/w^2)$ as
described by \citet{lfl+06}.  For those pulsars whose positions are
not accurately determined from timing measurements, this correction is
not made, and the tabulated flux densities and corresponding
luminosities are lower limits.

\section{Discussion}\label{sec:discssion}

\subsection{Derived parameters}

Various derived parameters for the new pulsars are presented in
Table~\ref{tb:deriv}. For each pulsar, when known, we list the base-10
logarithms of the characteristic age, surface dipole magnetic field
strength and rate of loss of rotational energy.  The final columns
contain the pulsar distances and height above the Galactic plane
computed from their DMs assuming the \citet{cl02} models for the
Galactic distribution of free electrons. As a group, these pulsars
have consistent properties to the larger samples published in the main
multibeam survey. We note that some of the distances (e.g.~15.3~kpc
for PSR~J1929+19) may be significantly overestimated due to uncertainties 
in the electron density model for this part of the Galactic plane.

\begin{table*} 
\caption{
Observed and derived parameters for the 18 pulsars presented in this
paper. Listed the Galactic longitude, $l$, Galactic latitude, $b$
(both in degrees), DM (cm$^{-3}$~pc), pulse width at 50\% and 10\% of
the peak and the equivalent width (respectively $W_{50}$, $W_{10}$ and
$W_{\rm eq}$ all in ms), peak and mean 1400-MHz flux density (mJy),
the base-10 logarithms of characteristic age (yr), the surface dipole
magnetic field strength (G), the loss in rotational energy
(erg~s$^{-1}$), the DM-derived distance using the Cordes \& Lazio
(2001) model, $D$ (kpc), the height with respect to the Galactic
plane, $z$ (pc) and the 1400-MHz pseudoluminosity (mJy~kpc$^2$).}
\label{tb:deriv}
\begin{center}\begin{footnotesize}
\begin{tabular}{lrrrrrrrrrrrrrr}
\hline 
\multicolumn{1}{c}{PSR J} & 
\multicolumn{1}{c}{$l$} & 
\multicolumn{1}{c}{$b$} & 
\multicolumn{1}{c}{DM} & 
\multicolumn{1}{c}{$W_{50}$} & 
\multicolumn{1}{c}{$W_{10}$} & 
\multicolumn{1}{c}{$W_{\rm eq}$} & 
\multicolumn{1}{c}{$S_{\rm peak}$} & 
\multicolumn{1}{c}{$S_{\rm mean}$} & 
\multicolumn{1}{c}{$\log[\tau_c]$} & 
\multicolumn{1}{c}{$\log[B]$} & 
\multicolumn{1}{c}{$\log[\dot{E}]$} & 
\multicolumn{1}{c}{$D$} & 
\multicolumn{1}{c}{$z$} &
\multicolumn{1}{c}{$L_{1400}$} \\
\hline 
1919+1645&51.0& 1.6 & 208(9) & 14.5 & 35.0 & 13.7 & 6.4 & 0.16 & 7.6 & 11.5 & 31.7 & 6.8 & 200 & 7.4  \\
1921+1544&50.4& 0.6 & 385(2) &  6.3 & 13.7 &  0.4 & 4.7 & 0.01 & 6.4 & 11.6 & 34.1 &13.1 & 140 & 1.7  \\
1922+1733&52.1& 1.2 & 238(4) &  9.4 & 16.7 &  8.3 & 8.7 & 0.31 & 5.4 & 12.3 & 34.6 & 7.4 & 160 & 17  \\
1924+1639&51.4& 0.6 & 208(3) &  7.2 & 10.9 &  7.6 &10.6 & 0.51 & 6.0 & 11.8 & 34.4 & 6.6 &  65 & 22  \\
1925+19  &53.7& 1.4 & 328(16)&  --- &130.2 & 55.5 &$>9$&$>0.5$&     &      &      & 9.8 & 250 & $>48$\\ \\
1926+2016&54.9& 1.8 & 247(5) &  9.7 & 18.0 & 13.1 &14.6 & 0.64 & 6.1 & 12.0 & 33.7 & 8.1 & 260 &  42 \\
1928+1923&54.3& 1.0 & 476(14)& 30.9 & 71.2 & 30.7 &18.6 & 0.70 & 6.3 & 12.4 & 32.7 &14.1 & 480 &  140 \\
1929+16  &51.7&--0.7& 12(9)  & 20.6 & 32.9 & 24.4 &$>7$&$>0.3$&     &      &      & 1.3 &--14 &   $>0.5$\\
1929+19  &54.2& 0.6 & 527(6) & 38.4 & 69.6 &  8.1 &$>3$&$>0.06$&    &      &      &15.3 & 180 &   $>14$\\
1929+1955&54.9& 1.0 & 281(4) & 10.2 & 16.4 &  7.1 &12.0  & 0.33 & 6.2 & 11.9 & 33.7 & 8.8 & 160 & 26\\ \\
1929+2121&56.1& 1.8 & 66(12) & 15.0 & 33.0 & 10.3 &16.2  & 0.23  & 6.7 & 12.1 & 32.3 & 3.4 & 100 &2.7   \\
1930+17  &52.8&--0.5& 201(27)& 68.6 & 93.4 &111.0 & $>3$  & $>0.2$  &     &      &      & 6.4 &--37 & $>8$  \\
1931+1952&55.1& 0.5 & 441(8) & 13.0 & 22.4 & 16.9 & 9.9 & 0.33 & 7.9 & 11.4 & 31.5 &12.6 & 100 & 52  \\
1935+1726&53.4&--1.4& 61.6(1)&  0.8 &  1.7 &  0.6 & 4.6 & 0.68 &     &      &      & 3.2 &--76 & 7  \\
1935+2025&56.1&--0.1& 182(1) &3.7,2.4&6.7,4.5& 4.9& 13.0 & 0.79 & 4.3 & 12.3 & 36.7 & 6.2 & --6 & 30\\ \\
1936+21  &56.8& 0.2 & 264(11)& 19.2 & 31.2 & 30.3 & $>2.2$ & $>0.03$ &     &      &      & 8.4 &  38 &$>2$   \\
1938+20  &56.1&--0.7& 306(12)& 47.9 & 79.1 & 50.4 & $>2.5$ & $>0.2$  &     &      &      & 9.3 &--110& $>17$  \\
1938+2213&57.9& 0.3 & 91(3)  &  6.8 & 15.1 &  6.6 & 14.8 & 0.59 & 4.8 & 12.4 & 35.5 & 4.1 &  22 & 10  \\
\hline 
\end{tabular}
\label{tb:parms}
\end{footnotesize}\end{center}\end{table*}

Also listed in Table~\ref{tb:parms} are the flux densities and the
corresponding pseudoluminosity estimates (i.e.~flux times distance
squared), or lower limits thereof, as appropriate. While these are
generally weak sources, with 1400-MHz flux densities as low as 10~$\mu$Jy,
because of their large distances, the pulsars'
radio luminosities are not unusually low when compared with the
rest of the known population.

\subsection{PSR~J1935+1726} \label{sec:1935+1726}

This 4.2~ms pulsar is one of two millisecond pulsars detected from an
analysis of the survey observations so far. The other detection was
the original millisecond pulsar, B1937+21 \citep{bkh+82}. While the
timing precision we have been able to achieve from PSR~J1935+1726
($\sim$~50~$\mu$s RMS) will mean that it is unlikely to warrant
consideration into pulsar timing arrays \citep[e.g.,][]{dfg+13}, it
does permit a precise determination of the orbital parameters. From
these data provided in Table~\ref{tb:binary}, the Keplerian mass
function for this system is $4.26 \times 10^{-3}$~M$_{\odot}$. This is
a typical value for millisecond pulsar binaries in which the companion
star is expected on evolutionary grounds to be a low-mass white dwarf
\citep[see, e.g.,][]{lor08}.  Assuming a millisecond pulsar mass of
1.5~M$_{\odot}$ \citep{opns12}, and an orbital inclination angle of
$90^{\circ}$, we infer the companion star's mass to be
$>0.23$~M$_{\odot}$. The orbital eccentricity for this system ($1.76
\times 10^{-4}$) listed in Table~\ref{tb:binary} is entirely
consistent with the value predicted from the orbital period by the
fluctuation--dissipation theory developed by \citet{phi92}. This has
been noted for many other \citep[but not all,][]{bkl+13} millisecond
pulsar binaries, suggesting that PSR~J1935+1736 has followed a standard 
low-mass binary evolution to evolve to its presently observed state.

\subsection{PSR~J1935+2025} \label{sec:1935+2025}

This 80.1~ms pulsar has the lowest characteristic age of all the new
discoveries ($\tau_c \approx 21,000$~yr), and is the only pulsar in
which an interpulse is clearly detectable (Fig.~\ref{fg:prf}).  No
known supernova remnant, X-ray or gamma-ray source exists at the
pulsar's position.  
There is a significant amount of unmodeled behaviour still
present in the timing residuals.
This can be largely removed by fitting for a second frequency
derivative of $\ddot{\nu} = (2.93 \pm 0.06) \times
10^{-22}$~s$^{-3}$. When interpreted in terms of the standard
power-law spindown model where $\dot{\nu} \propto \nu^n$, for a
braking index $n$ and spin frequency $\nu$ \citep[e.g.,][]{lk05}, the
implied braking index $n=\nu \ddot{\nu}/(\dot{\nu})^2=40.8$ is well
outside the range expected for long-term spin behaviour
\citep[e.g.,][]{jg99} and most likely suggests the presence of
significant low-frequency noise in this pulsar.

\begin{figure} 
\centerline{\psfig{file=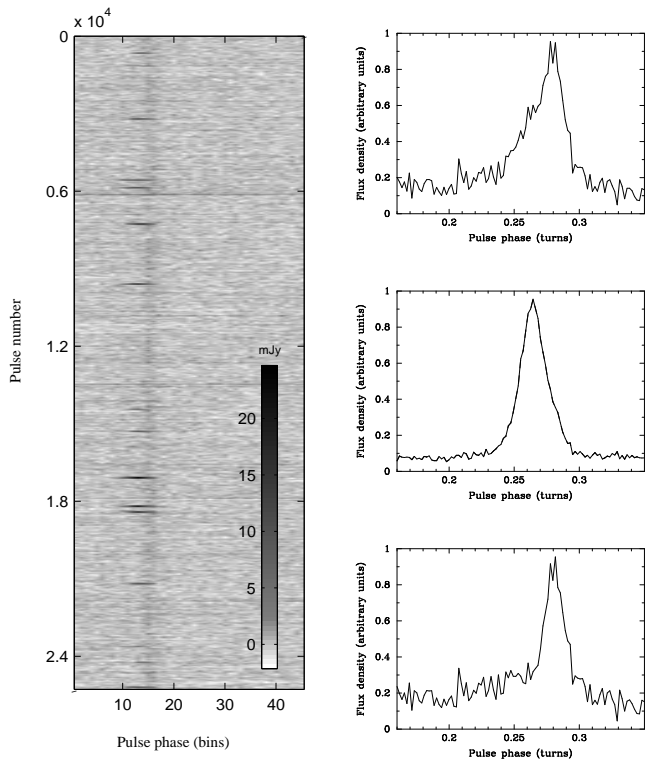,width=85mm}} 
\caption{Left:observing time versus pulse phase in bins for one of the survey
pointings for PSR~J1938+2213 showing bursting events during the 70-min
observation. Right: integrated pulse profiles showing 20\% of the
period around the main pulse for (top) the entire observation, (middle) the
bright bursts minus the average faint emission, (bottom) the faint emission
only. The millijansky scale was established using the radiometer
noise as described in Section 2.}
\label{fg:1938}
\end{figure}

\subsection{PSR~J1938+2213} \label{sec:1938+2213}

This 166-ms pulsar was discovered by Chandler (2003) in a 430-MHz
drift scan survey with the Arecibo telescope. As noted by Chandler,
this source is unusual in that it exhibits short timescale bursting
behaviour.  This emission is seen clearly in one of the survey
pointings shown in Fig.~\ref{fg:1938}. Each pulse phase bin shown in
this image has been smoothed using a boxcar of width 64 pulses running
vertically to enhance the average emission seen around pulse phase bin
number 15. A total of 12 bursting events, lasting typically 20--25
pulse periods, can be seen in this image. As the integrated profiles
in the inset to this figure show, the bursting events are clearly
offset from the main profile in pulse phase. Thus the emission process
in this pulsar appears to consist of a steady emission mode superposed
by these brighter bursts. Our understanding of these bursting pulsars
is currently limited by studies of only a few sources
\citep{now92,lwf+04,slr13}. Further examples of this phenomenon, as
well as more sensitive observations, are clearly required to improve
our understanding of the complex emission physics of neutron stars.

\section{Conclusions}\label{sec:conclusions}

We have carried out a deep survey of the northern Galactic plane
using the Parkes radio telescope and have discovered 17 new pulsars.
Among the new discoveries is the binary millisecond pulsar PSR~J1935+1726
and the young 80.1-ms pulsar PSR~J1935+2025. Our follow up timing observations
have provided phase-connected timing solutions for 11 of these pulsars,
as well as a solution for the previously known bursting
pulsar PSR~J1938+2213. Our survey observation of this pulsar
reveal that the bursting emission is superposed over a fainter,
steadier, emission region which is offset in pulse phase from the bursts.

A novel feature of our survey was that we carried out two observations
of each sky position. All of the discoveries made here were visible in
both survey observations. Further processing of the data in future will
search for fainter candidates which might appear in both survey
pointings, but at a lower significance than would otherwise be
believable from an analysis of a single survey pointing. So far, our
processing has not accounted for the effects of any orbital motion
which may be significant during the 70-min pointings. An acceleration
search analysis will shortly be carried out and the results of this
search will be published elsewhere.

\section*{Acknowledgments} 

The Parkes radio telescope is part of the Australia Telescope which is
funded by the Commonwealth of Australia for operation as a National
Facility managed by CSIRO. The National Radio Astronomy Observatory is
a facility of the National Science Foundation operated under
cooperative agreement by Associated Universities, Inc.  This work made
use of the facilities of the ATNF Pulsar Catalogue, the SAO/NASA
Astrophysics Data System and the HEASARC archival data search
tool. DRL acknowledges support from the Royal Society as a University
Research Fellow during the early phases of this project. We thank
Matthew Bailes for making his computing facilities at Parkes available
for the initial search processing.  Computer resources used during the
later stages of this project were supported from a WV EPSCoR Challenge
Grant awarded to DRL and MAM. We thank Andrew Seymour for producing
Figure 2, and West Virginia University undergraduates Austin
Anuta-Darling and Adil Moghal for data processing assistance during
the course of this work. DRL and MAM acknowledge support from
Oxford Astrophysics while on sabbatical leave.

\bibliographystyle{mn2e}
\bibliography{journals,modrefs,psrrefs,crossrefs}

\begin{thebibliography}{22}
\expandafter\ifx\csname natexlab\endcsname\relax\def\natexlab#1{#1}\fi

\bibitem[{Backer {et~al}\mbox{.}(1982)Backer, Kulkarni, Heiles, Davis, \&
  Goss}]{bkh+82}
Backer D.~C., Kulkarni S.~R., Heiles C., Davis M.~M., Goss W.~M., 1982, Nature,
  300, 615

\bibitem[{{Burgay} {et~al}\mbox{.}(2013){Burgay}, {Keith}, {Lorimer},
  {Hassall}, {Lyne}, {Camilo}, {D'Amico}, {Hobbs}, {Kramer}, {Manchester},
  {McLaughlin}, {Possenti}, {Stairs}, \& {Stappers}}]{bkl+13}
{Burgay} M. {et~al.}, 2013, MNRAS, 429, 579

\bibitem[{{Camilo}(2004)}]{cam04}
{Camilo} F., 2004, in IAU Symposium, Vol. 218, Young Neutron Stars and Their
  Environments, {Camilo} F., {Gaensler} B.~M., eds., p.~97

\bibitem[{Chandler(2003)}]{cha03}
Chandler A.~M., 2003, PhD thesis, California Institute of Technology

\bibitem[{{Cordes} {et~al}\mbox{.}(2006){Cordes}, {Freire}, {Lorimer},
  {Camilo}, {Champion}, {Nice}, {Ramachandran}, {Hessels}, {Vlemmings}, {van
  Leeuwen}, {Ransom}, {Bhat}, {Arzoumanian}, {McLaughlin}, {Kaspi}, {Kasian},
  {Deneva}, {Reid}, {Chatterjee}, {Han}, {Backer}, {Stairs}, {Deshpande}, \&
  {Faucher-Gigu{\`e}re}}]{cfl+06}
{Cordes} J.~M. {et~al.}, 2006, ApJ, 637, 446

\bibitem[{{Cordes} \& {Lazio}(2002)}]{cl02}
{Cordes} J.~M., {Lazio} T.~J.~W., 2002, {astro-ph/0207156}

\bibitem[{{Demorest} {et~al}\mbox{.}(2013){Demorest}, {Ferdman}, {Gonzalez},
  {Nice}, {Ransom}, {Stairs}, {Arzoumanian}, {Brazier}, {Burke-Spolaor},
  {Chamberlin}, {Cordes}, {Ellis}, {Finn}, {Freire}, {Giampanis}, {Jenet},
  {Kaspi}, {Lazio}, {Lommen}, {McLaughlin}, {Palliyaguru}, {Perrodin},
  {Shannon}, {Siemens}, {Stinebring}, {Swiggum}, \& {Zhu}}]{dfg+13}
{Demorest} P.~B. {et~al.}, 2013, ApJ, 762, 94

\bibitem[{{DuPlain} {et~al}\mbox{.}(2008){DuPlain}, {Ransom}, {Demorest},
  {Brandt}, {Ford}, \& {Shelton}}]{drd+08}
{DuPlain} R., {Ransom} S., {Demorest} P., {Brandt} P., {Ford} J., {Shelton}
  A.~L., 2008, in Society of Photo-Optical Instrumentation Engineers (SPIE)
  Conference Series, Vol. 7019, Society of Photo-Optical Instrumentation
  Engineers (SPIE) Conference Series

\bibitem[{Johnston \& Galloway(1999)}]{jg99}
Johnston S., Galloway D., 1999, MNRAS, 306, L50

\bibitem[{{Kaplan} {et~al}\mbox{.}(2005){Kaplan}, {Escoffier}, {Lacasse},
  {O'Neil}, {Ford}, {Ransom}, {Anderson}, {Cordes}, {Lazio}, \&
  {Kulkarni}}]{kel+05}
{Kaplan} D.~L. {et~al.}, 2005, PASP, 117, 643

\bibitem[{Lange {et~al}\mbox{.}(2001)Lange, Camilo, Wex, Kramer, Backer, Lyne,
  \& Doroshenko}]{lcw+01}
Lange C., Camilo F., Wex N., Kramer M., Backer D., Lyne A., Doroshenko O.,
  2001, MNRAS, 326, 274

\bibitem[{{Lewandowski} {et~al}\mbox{.}(2004){Lewandowski}, {Wolszczan},
  {Feiler}, {Konacki}, \& {So{\l}tysi{\' n}ski}}]{lwf+04}
{Lewandowski} W., {Wolszczan} A., {Feiler} G., {Konacki} M., {So{\l}tysi{\'
  n}ski} T., 2004, ApJ, 600, 905

\bibitem[{Lorimer(2008)}]{lor08}
Lorimer D.~R., 2008, Living Reviews in Relativity, 11

\bibitem[{{Lorimer} {et~al}\mbox{.}(2006){Lorimer}, {Faulkner}, {Lyne},
  {Manchester}, {Kramer}, {McLaughlin}, {Hobbs}, {Possenti}, {Stairs},
  {Camilo}, {Burgay}, {D'Amico}, {Corongiu}, \& {Crawford}}]{lfl+06}
{Lorimer} D.~R. {et~al.}, 2006, MNRAS, 372, 777

\bibitem[{Lorimer \& Kramer(2005)}]{lk05}
Lorimer D.~R., Kramer M., 2005, {Handbook of Pulsar Astronomy}. Cambridge
  University Press

\bibitem[{Manchester {et~al}\mbox{.}(2001)Manchester, Lyne, Camilo, Bell,
  Kaspi, D'Amico, McKay, Crawford, Stairs, Possenti, Morris, \&
  Sheppard}]{mlc+01}
Manchester R.~N. {et~al.}, 2001, MNRAS, 328, 17

\bibitem[{McLaughlin {et~al}\mbox{.}(2006)McLaughlin, Lyne, Lorimer, Kramer,
  Faulkner, Manchester, Cordes, Possenti, Camilo, Hobbs, Stairs, D'Amico, \&
  O'Brien}]{mll+06}
McLaughlin M.~A. {et~al.}, 2006, Nature, 439, 817

\bibitem[{{Morris} {et~al}\mbox{.}(2002){Morris}, {Hobbs}, {Lyne}, {Stairs},
  {Camilo}, {Manchester}, {Possenti}, {Bell}, {Kaspi}, {Amico}, {McKay},
  {Crawford}, \& {Kramer}}]{mhl+02}
{Morris} D.~J. {et~al.}, 2002, MNRAS, 335, 275

\bibitem[{{Nowakowski}(1992)}]{now92}
{Nowakowski} L.~A., 1992, in IAU Colloq. 128: Magnetospheric Structure and
  Emission Mechanics of Radio Pulsars, {Hankins} T.~H., {Rankin} J.~M., {Gil}
  J.~A., eds., p. 280

\bibitem[{{{\"O}zel} {et~al}\mbox{.}(2012){{\"O}zel}, {Psaltis}, {Narayan}, \&
  {Santos Villarreal}}]{opns12}
{{\"O}zel} F., {Psaltis} D., {Narayan} R., {Santos Villarreal} A., 2012, ApJ,
  757, 55

\bibitem[{Phinney(1992)}]{phi92}
Phinney E.~S., 1992, Philos. Trans. Roy. Soc. London A, 341, 39

\bibitem[{Seymour {et~al}\mbox{.}(2013)Seymour, Lorimer, \& Ridley}]{slr13}
Seymour A., Lorimer D.~R., Ridley J.~P., 2013, MNRAS, submitted

\end{thebibliography}
\end{document}